\documentclass[twocolumn,showpacs,showkeys,prd,floatfix,apsrev.bst]{revtex4}
\usepackage{ifthen,graphicx}
\usepackage{bm}
\usepackage{amsmath,amssymb}

\begin{document}
\title{Broken symmetry can yield a positive effective G  \\
in conformal gravity}

\date{\today}

\author{Peter~R.~Phillips}
\affiliation{Department of Physics, Washington University, St.~Louis,
MO 63130 }
\email{prp@wuphys.wustl.edu}

\begin{abstract}
We modify the action of Mannheim's conformally invariant model by changing
the sign of two coefficients. This breaks conformal symmetry, but results in
a cosmology that has a positive effective G and at the same time retains one
of the main advantages of the Mannheim model, a possible solution of the
cosmological constant problem.
\end{abstract}

\pacs{04.40.Nr, 04.50.Kd}
\keywords{gravitation; cosmology: theory}

\maketitle


\section{Introduction}
\label{sec:intro}

In a series of articles following his seminal paper of 1989 \cite{mann04},
Mannheim has developed an alternate theory of gravitation called conformal
gravity. We will be concerned here with the application of this theory
to cosmology; see the review \cite{mann6}, referred to as PM from now on.
The field equations are (PM, equation (186)):
\begin{equation}
4 \alpha_g W^{\mu \nu} = T^{\mu \nu}
\label{eq:mannfield}
\end{equation}
where $W^{\mu \nu}$ is the Weyl conformal tensor (PM, equations (185),
(107) and (108)), and $\alpha_g $ is a dimensionless coupling constant.
$T^{\mu \nu}$ is the energy-momentum tensor, defined in such a way that the
field equations are invariant under conformal transformations (PM, equation
(176)):
\begin{equation}
g_{\mu \nu} \rightarrow e^{2 \alpha(x)} g_{\mu \nu}
\label{eq:contrans}
\end{equation}

Mannheim includes a fermion field $\psi$ and a scalar
field $S$, and writes the action as (PM, equation (61)):
\begin{eqnarray}
I_M & = & - \int {\rm d}^4 x (-g)^{1/2}  \left\{
\frac{1}{2} S^{;\mu} S_{;\mu}
- \frac{1}{12} S^2 R^{\mu}_{\;\;\mu} \right. \nonumber \\
  & & \hspace{-20pt} \left. \rule{0pt}{14pt} {} + \lambda S^4
+ i \overline{\psi} \gamma^{\mu} (x) \left[ \partial_{\mu}
+ \Gamma_{\mu} (x) \right] \psi - h S \overline{\psi}\psi \right\}
\label{eq:mannaction}
\end{eqnarray}
We use the notation of Mannheim: metric signature is $(- + + +)$, 
$g = + {\rm Det} (g_{\mu \nu})$, and the curvature tensors are defined as in
Weinberg \cite{wein2}. We set $c=1$ except when comparing results to those
of general relativity. $\lambda$ and $h$ are dimensionless coupling
constants. The factor $1/12$ in the $R^{\mu}_{\;\;\mu}$ term is necessary
for the resulting field equations to be conformally invariant (see, for
example, Birrell and Davies \cite{birdav}). We shall not be concerned with
fermion fields in this paper. Omitting them, this action yields the
field equation for $S$ (PM, equation (63)):
\begin{equation}
S^{;\mu}_{\;\;\mu} + \frac{1}{6} S R^{\mu}_{\;\;\mu} - 4 \lambda S^3 = 0
\label{eq:mannS}
\end{equation}
and the expression for $T_{\mu \nu}$ (PM, equation (64)):

\begin{eqnarray}
T_{\mu \nu} & = & \frac{2}{3} S_{;\mu} S_{;\nu}
- \frac{1}{6} g_{\mu \nu} S^{;\alpha}_{\;\;;\alpha}
- \frac{1}{3} S S_{;\mu ;\nu}
 + \frac{1}{3} g_{\mu \nu} S S^{;\alpha}_{\;\;;\alpha}
\nonumber \\
  & & {} - \frac{1}{6} S^2 \left( R_{\mu \nu}
-  \frac{1}{2} g_{\mu \nu} R^{\alpha}_{\;\;\alpha} \right)
- g_{\mu \nu} \lambda S^4
\label{eq:mannT}
\end{eqnarray}

\section{Application to cosmology}

We assume that on the largest scales our Universe can be adequately
described as a Friedmann-Robertson-Walker (FRW) space. Adapting the field
equations to such a space results in the Mannheim model, the main
features of which are recapitulated in this section.

The metric tensor for  FRW space can be written
\begin{equation}
g_{\mu \nu} = {\rm Diag } \left[ -1, \frac{R^2 (t)}{1 - k r^2 },
R^2 (t) r^2, R^2 (t) r^2 \sin^2 \theta \right]
\label{eq:FRWmetric}
\end{equation}
where $R(t)$ is the expansion factor and $k$ can take the values $1$, $0$,
or $-1$. Mannheim opts for $k=-1$ in his model. From this metric tensor we
can derive an expression for $R^{\alpha}_{\;\;\alpha}$:
\begin{equation}
R^{\alpha}_{\;\;\alpha} = -\frac{6}{R^2 } \left(
k + \dot{R}^2 + R \ddot{R} \right)
\label{eq:Ricci}
\end{equation}

Since a FRW space is conformally flat, the Weyl tensor vanishes, and the
field equations (\ref{eq:mannfield}) reduce to (PM, equation (219)):
\begin{equation}
T^{\mu\nu} = 0
\label{eq:mannfield2}
\end{equation}

Mannheim assumes a conformal transformation is made that reduces the field
$S$ to a constant value, $S_0 $. The field equation (\ref{eq:mannS}) then
requires
\begin{equation}
S^{2}_{0} = \frac{1}{24 \lambda} R^{\alpha}_{\;\;\alpha}
\label{eq:mannS2}
\end{equation}
and (\ref{eq:mannfield2}) becomes, using (\ref{eq:mannT}),
\begin{equation}
T^{\mu \nu}_{\rm kin} - \frac{1}{6} S^{2}_{0} \left( R^{\mu \nu}
- \frac{1}{2} g^{\mu \nu} R^{\alpha}_{\;\;\alpha} \right)
- \lambda S^{4}_{0} g^{\mu \nu} = 0
\label{eq:mannfield3}
\end{equation}
where $T^{\mu \nu}_{\rm kin}$ is the energy-momentum tensor of any matter
fields. (In this paper, since we neglect fermion fields, we assume this
tensor derives entirely from electromagnetic radiation.)

Writing (\ref{eq:mannfield3}) in the form
\begin{equation}
R^{\mu \nu} - \frac{1}{2} g^{\mu \nu} R^{\alpha}_{\\;\alpha} =
\frac{6}{S^{2}_{0}} \left( T^{\mu \nu}_{\rm kin}
- \lambda S^{4}_{0} g^{\mu \nu} \right)
\label{eq:mannfield4}
\end{equation}
we see that the field equations are just those of general relativity, but
with an effective gravitational constant that is {\em negative}
(PM, equation (224)):
\begin{equation}
G_{\rm eff} = - \frac{3 c^3 }{4 \pi S^{2}_{0}}
\label{eq:mannGeff}
\end{equation}
and a cosmological constant
\begin{equation}
\Lambda = \lambda S^{4}_{0}
\label{eq:manncoscon}
\end{equation}

Mannheim chooses $\lambda < 0$, and assumes $T^{\mu \nu}_{\rm kin}$ consists
in radiation with a density $\rho_m = A/R^4 (t)$. He then solves the
Friedmann equation derived from the field equations (\ref{eq:mannfield4}) to
get (PM, equation (230)):
\begin{equation}
R^2 (t) = - \frac{k(\beta-1)}{2 \alpha} 
- \frac{k \beta \sinh^2 \left( \alpha^{1/2} ct \right)}{\alpha}
\label{eq:mannR}
\end{equation}
where
\begin{eqnarray}
\alpha c^2 & = & -2 \lambda S^{2}_{0} =
\frac{8 \pi G_{\rm eff} \Lambda}{3c}
\label{eq:mannalpha} \\
\beta & = & \left( 1 - \frac{16 A \lambda}{k^2 c} \right)^{1/2}
\label{eq:mannbeta}
\end{eqnarray}
Note that, since $\lambda < 0$, $\alpha$ is positive and $\beta > 1$.

According to (\ref{eq:mannR}), $R(t)$ does not go to zero as
$t \rightarrow 0$, but to the finite value
$\left[(\beta-1)/(2 \alpha) \right]^{1/2}$.

Mannheim argues (PM, page 426) that this solution may hold the key to the
cosmological constant problem, since the quantity
\begin{equation}
\overline{\Omega}_{\Lambda} (t) \equiv
\frac{8 \pi G_{\rm eff} \Lambda}{3 c H^{2} (t)}
\label{eq:mannOmegaLambda}
\end{equation}
goes like $\tanh^{2} \left(\alpha^{1/2} c t \right)$ for large $t$, and
hence is bounded from below no matter how large $\Lambda$ may be.

\section{Breaking the conformal symmetry}

The Mannheim model has some attractive features, but a basic drawback in
that the effective gravitational constant, $G_{\rm eff}$, is negative. One
can argue that this negative value is only apparent on the largest scales,
but it seems quite unlikely that, for example, the observations of the
cosmic microwave background (CMB) can be reconciled with such a
$G_{\rm eff}$.

In this section we introduce ideas that are familiar from inflation and
particle physics, and try to modify Mannheim's procedure so as to achieve a
$G_{\rm eff}$ that is positive. The simplest solution to Mannheim's field
equations is Minkowski space, where $k = 0$ and $\lambda = 0$. Since the
space is flat, $R^{\alpha}_{\;\;\alpha}$ is also zero, and (\ref{eq:mannS})
can be satisfied with an arbitrary value of $S_0 $.

Let us suppose that a space of this sort is metastable against the formation
of bubbles of lower symmetry. Once a bubble forms it will expand at the
velocity of light, and the geometry inside the bubble is of FRW type, with
$k=-1$ (the idea of such bubble formation goes back to 1980 \cite{colelucc},
and has been discussed many times since, for example in \cite{bgt}).

We will assume the field equations are still given by (\ref{eq:mannfield}),
but the action (\ref{eq:mannaction}) is replaced by
\begin{eqnarray}
I_M & = & - \int {\rm d}^4 x (-g)^{1/2}  \left\{
\frac{1}{2} S^{;\mu} S_{;\mu}
- \frac{\mu}{12} S^2 R^{\mu}_{\;\;\mu} \right. \nonumber \\
  & & \hspace{-20pt} \left. \rule{0pt}{14pt} {} + \lambda S^4
+ i \overline{\psi} \gamma^{\mu} (x) \left[ \partial_{\mu}
+ \Gamma_{\mu} (x) \right] \psi - h S \overline{\psi}\psi \right\}
\label{eq:prpaction}
\end{eqnarray}
where the coefficient $\mu$ in the $R^{\mu}_{\;\;\mu}$ term need not have
the value unity, so general conformal invariance is broken. We still, of
course, have invariance under the restricted conformal transformation
\begin{equation}
g_{\mu \nu} \rightarrow e^{2 \alpha} g_{\mu \nu}
\label{eq:contrans2}
\end{equation}
with $\alpha$ a constant, and no longer a general function of $x^{\mu}$.

We will not concern ourselves with details of the bubble formation, and
the reader may prefer simply to start from the field equations
(\ref{eq:mannfield}) and the action (\ref{eq:prpaction}).

Omitting fermion fields, the field equation for $S$ becomes (see
(\ref{eq:mannS})):
\begin{equation}
S^{;\mu}_{\;\;\mu} + \frac{\mu}{6} S R^{\mu}_{\;\;\mu} - 4 \lambda S^3 = 0
\label{eq:prpS}
\end{equation}
and the expression for $T_{\mu \nu}$ (see (\ref{eq:mannT})):
\begin{eqnarray}
T_{\mu \nu} & = & \frac{2}{3} S_{;\mu} S_{;\nu}
- \frac{1}{6} g_{\mu \nu} S^{;\alpha}_{\;\;;\alpha}
- \frac{1}{3} S S_{;\mu ;\nu}
 + \frac{1}{3} g_{\mu \nu} S S^{;\alpha}_{\;\;;\alpha}
\nonumber \\
  & & {} - \frac{\mu}{6} S^2 \left( R_{\mu \nu}
-  \frac{1}{2} g_{\mu \nu} R^{\alpha}_{\;\;\alpha} \right)
- g_{\mu \nu} \lambda S^4
\label{eq:prpT}
\end{eqnarray}

In a FRW space we can assume $S$ is a function of $t$ alone, and the
equation of motion becomes
\begin{equation}
\ddot{S} R^2 + 3 \dot{S} R \dot{R}
+ \mu kS + \mu S \dot{R}^2 + \mu SR \ddot{R} + 4 \lambda S^3 R^2 = 0
\label{eq:M6_63_mod}
\end{equation}

Two independent equations result from $T_{\mu \nu} = 0$;
we take them to be the (0 0) and (1 1) components:
\begin{equation}
R^2 \dot{S}^2 + 2 S \dot{S} R \dot{R}
+ \mu k S^2 + \mu S^2 \dot{R}^2 + 2 \lambda S^4 R^2 = 0
\label{eq:M6_64_00_mod}
\end{equation}
\begin{eqnarray}
4 S \dot{S} R \dot{R} - R^2 \dot{S}^2 + 2 R^2 S \ddot{S} & & {}
\nonumber \\
{} + \mu k S^2 + \mu S^2 \dot{R}^2 + 2 \mu S^2 R \ddot{R}
+ 6 \lambda S^4 R^2 & = & 0
\label{eq:M6_64_11_mod}
\end{eqnarray}

Now (\ref{eq:M6_64_00_mod}) can be derived from (\ref{eq:M6_63_mod}) and
(\ref{eq:M6_64_11_mod}), so we actually only have two independent equations
for $R$ and $S$, which we can take to be (\ref{eq:M6_63_mod}) and
(\ref{eq:M6_64_11_mod}). But this turns out to be a singular pair. If we were
to try to integrate them numerically, we would have to rearrange them into
the form
\begin{eqnarray}
\ddot{R} & = & f_1 (R, S, \dot{R}, \dot{S}) \\
\ddot{S} & = & f_2 (R, S, \dot{R}, \dot{S}) 
\end{eqnarray}
but we find easily that such a rearrangement is impossible.

We can, however, solve these equations by simply choosing for $S$ the
solution $S = S_0 = {\rm constant}$. Having done that, we can solve for $R$,
then check back to ensure that our solution is consistent with our choice
for $S$. We do not need to invoke a conformal transformation to do this; we
just have to specify $S$ before we can solve for $R$. (Is
$S = {\rm constant}$ the only consistent choice? We do not know.)

The calculation now follows the same pattern as Mannheim's. We get a
Friedmann equation of the same form, except for the inclusion of a factor
$\mu$, so we have to solve
\begin{equation}
R^{\mu \nu} - \frac{1}{2} g^{\mu \nu} R^{\alpha}_{\\;\alpha} =
\frac{6}{\mu S^{2}_{0}} \left( T^{\mu \nu}_{\rm kin}
- \lambda S^{4}_{0} g^{\mu \nu} \right)
\label{eq:prpfield}
\end{equation}
which gives
\begin{equation}
G_{\rm eff} = - \frac{3 c^3 }{4 \pi \mu S^{2}_{0}}
\label{eq:prpGeff}
\end{equation}
This will be positive if we choose $\mu < 0$. The reversal of sign of the
$S^2$ term in the action (\ref{eq:prpaction}) has been noted before in
connection with $\phi^4$ field theory, which the Mannheim model resembles;
see, for example, \cite{kl} -- \cite{narpad}.

We will also reverse the sign of $\lambda$, so it is now positive. A few
words need to be said about this choice also. As noted, the field equation
(\ref{eq:mannS}) for $S$ resembles that of the scalar field, $\phi$, in
conventional $\phi^4 $ theory. In particular, $\lambda > 0$ is the
``right-sign'' choice, as illustrated in \cite{kl} -- \cite{narpad}. This
insures a spectrum that is bounded below. Choosing $\lambda < 0$, as
Mannheim does, entails treating the theory as a PT symmetric one
\cite{cmb3}.

Introducing radiation with energy density $\rho = A/R^4 (t)$,
the Friedmann equation becomes (PM, equations (225) and (226)):
\begin{equation}
\dot{R}^2 = \frac{8 \pi  G_{\rm eff} A}{3 c^2 R^2 }
- kc^2 + \frac{8 \pi G_{\rm eff} \Lambda R^2 }{3c}
\label{eq:prpFried}
\end{equation}
Here $\Lambda = \lambda S^{4}_{0}$ and $k = -1$, as before.

Provided $A$ is not too large, this has an exact solution that can be
written in a notation analogous to that of Mannheim:
\begin{eqnarray}
R^2 (t) & = & \frac{(\beta - 1)}{2 \alpha}
+ \frac{\beta}{\alpha} \sinh^2 \left[ \alpha^{1/2} c \,(t + \gamma )\right]
\label{eq:prpR} \\
\alpha & = & - \frac{2 \lambda S^2_0 }{ \mu }  > 0 \;\;
{\rm for} \;\; \mu < 0\;\;{\rm and}\;\;\lambda > 0
\label{eq:prpalpha} \\
\beta & = & \left[ 1 - \frac{16 A \lambda}{\mu^2 c } \right]^{1/2} < 1
\;\;{\rm for}\;\;\lambda > 0
\label{eq:prpbeta} \\
\gamma & = & \frac{1}{\alpha^{1/2} c}\,{\rm arcsinh}
\left[ \left( \frac{1-\beta}{2 \beta} \right)^{1/2} \right]
\label{eq:prpgamma}
\end{eqnarray}
\hspace{2in}

With this choice for $\gamma$, $R(t) \rightarrow 0$ as $t \rightarrow 0$.
From (\ref{eq:prpbeta}) we obtain the upper limit for $A$ for this analytic
solution to be valid:
\begin{equation}
A <= \frac{\mu^2 c}{16}
\label{eq:prpA}
\end{equation}

Using (\ref{eq:Ricci}) we can show
\begin{equation}
R^{\alpha}_{\;\;\alpha} = \frac{24 \lambda S^2_0 }{\mu }
\end{equation}
so that that (\ref{eq:prpS}) is satisfied when $S = S_0$, and we have a
consistent solution to our field equations.

\section{Comparison with the Mannheim model}

Points of agreement: the two models coincide for large $t$, when the
$\Lambda$ term dominates. Further, if Mannheim is correct in claiming that
his model solves the cosmological constant problem, then the same can be
said for the present model, since $\overline{\Omega}_{\Lambda}$ has a
similar behavior. Both models have $k = -1$.

Points of difference: the main one, of course, is that the present model has
a positive $G_{\rm eff}$. As a consequence, $R(t)$ tends to zero, rather
than a non-zero value, as $t \rightarrow 0$.

\section{Conclusion}

By breaking the conformal symmetry in Mannheim's model, we have constructed
a new model with a positive $G_{\rm eff}$. This model has an attractive
feature of the Mannheim model (possible solution of the cosmological
constant probelm), and at the same time is sufficiently similar
to conventional general relativity that it may prove capable of explaining
cosmological observations.

If we assume the field equations (\ref{eq:mannfield}) remain the same, with
the new $T^{\mu \nu}$, then one of the main achievements of conformal
gravity, the interpretation of galactic rotation curves, is unaltered.
Cosmology has nothing to say about the coupling constant $\alpha_g$, since
the Weyl tensor vanishes in a FRW space. We note, however, that if we are to
obtain the pleasing behavior of the gravitational potential described in
\cite{prpmnras}, $\alpha_g$ must be taken to be positive.

\begin{acknowledgments}
Calculations in this paper have been aided by use of the program GRTensorII.
\end{acknowledgments}


\end{document}